%% file: main.tex
\documentclass[sigconf]{acmart}

\usepackage{natbib}
\usepackage{multirow}
\usepackage{graphicx}
\usepackage{multicol}
\usepackage{todonotes}
\usepackage{subcaption}
\usepackage{listings}
\usepackage{varwidth}
\usepackage[ruled,vlined]{algorithm2e}

\newcommand{\bumpup}{\vspace*{-.5cm}}

\setcopyright{acmlicensed}
\copyrightyear{2026}
\acmYear{2026}
\setcopyright{rightsretained}
\acmConference[UMAP’26]{UMAP’26 Conference}{June, 2026}{Gothenburg}
\acmDOI{}
\acmISBN{}

\begin{document}

\title{Multistakeholder Impacts of Profile Portability in a Recommender Ecosystem}

\author{Anas Buhayh}
\email{anas.buhayh@colorado.edu}
\orcid{0009-0009-7987-7967}
\affiliation{%
  \institution{Department of Information Science, University of Colorado, Boulder}
  \city{Boulder}
  \state{Colorado}
  \country{USA}
  \postcode{80309}
}

\author{Elizabeth McKinnie}
\email{elizabeth.mckinnie@colorado.edu}
\orcid{0009-0002-8721-5700}
\affiliation{%
  \institution{Department of Information Science, University of Colorado, Boulder}
  \city{Boulder}
  \state{Colorado}
  \country{USA}
  \postcode{80309}
}

\author{Clement Canel}
\email{clement.canel@colorado.edu}
\orcid{0009-0003-1982-8174}
\affiliation{%
  \institution{Department of Computer Science, University of Colorado, Boulder}
  \city{Boulder}
  \state{Colorado}
  \country{USA}
  \postcode{80309}
}

\author{Robin Burke}
\email{robin.burke@colorado.edu}
\orcid{0000-0001-5766-6434}
\affiliation{%
  \institution{Department of Information Science, University of Colorado, Boulder}
  \city{Boulder}
  \state{Colorado}
  \country{USA}
  \postcode{80309}
}

\begin{abstract}

Optimizing outcomes for multiple stakeholders in recommender systems has historically focused on algorithmic interventions, such as developing multi-objective models or re-ranking results from existing algorithms. However, structural changes to the recommendation ecosystem itself remain understudied. This paper explores the implications of algorithmic pluralism (also known as "middleware" in the governance literature), in which recommendation algorithms are decoupled from platforms, enabling users to select their preferred algorithm. Prior simulation work demonstrates that algorithmic choice benefits niche consumers and providers. Yet this approach raises critical questions about user modeling in the context of data portability: when users switch algorithms, what happens to their data? Noting that multiple data portability regulations have emerged to strengthen user data ownership and control. We examine how such policies affect user models and stakeholders' outcomes in recommendation setting. Our findings reveal that data portability scenarios produce varying effects on user utility across different recommendation algorithms. We highlight key policy considerations and implications for designing equitable recommendation ecosystems.

\end{abstract}



\maketitle

\input{1_intro_related}
\input{2_smores}
\input{3_methodology}
\input{4_results}
\input{5_conclusion}

\bibliographystyle{ACM-Reference-Format}
\bibliography{references}

\end{document}

%% file: 1_intro_related.tex
\section{Introduction}

There is substantial literature on multistakeholder recommender systems, outlined in \cite{Abdollahpouri2020}. Various interventions have been designed to balance system benefits across stakeholders. We see two main types of interventions: model-based (incorporating multistakeholder objectives into the recommendation model) and re-ranking (applying post-hoc adjustments to recommendation lists). An alternative point of intervention is at the ecosystem level: the multistakeholder properties of an individual recommender system may be less critical if a variety of recommendation algorithms with different properties are available. Market-based recommendation ecosystems have been proposed under various names: \textit{friendly neighborhood algorithm store}~\cite{rajendra2023three}, \textit{middleware}~\cite{fukuyama2021save,hogg2024shapingfuturesocialmedia}, or \textit{algorithmic pluralism}~\cite{verhulst2023steering}. As in other economic sectors, we might expect algorithm designers to specialize in serving particular audiences, and stakeholder outcomes — both consumer-side and provider-side — may be enhanced over solutions that rely on a single recommender system to be all things to all people. 

Using a simulation-based approach, \citet{buhayh2024decoupled, buhayh2025simulating} explored the dynamics of a recommender system marketplace with two algorithm providers: one for a generic audience and one focused on a particular content niche. These studies found that a multistakeholder recommender ecosystem created fairer outcomes for users with distinct tastes outside of the mainstream, and those who produced content catering to niche tastes were able to achieve better outcomes with minimal loss to the utility experienced by other consumers and providers. 

In this work, we extend this line of research by investigating how regulatory choices, specifically around user profile management, impact a multi-algorithm recommender ecosystem and the outcomes for niche consumers and providers. In particular, we study the question of profile portability: to what extent are user profiles shared (or not) among algorithm operators when users move from one algorithm to another? 

We explore two dimensions of portability. The first is \textit{exclusivity}, the extent to which a user's profile data remains tied to a given algorithm and does not follow the user when switching to a new recommender. Since algorithm operators compete, they may be unwilling to share user data with rivals, creating significant `lock-in' as users may hesitate to switch if their data cannot move with them. Article 20 of the GDPR \cite{wp2017dataportability} grants users the right to download or transfer their data, which may be interpreted as requiring non-exclusivity.

The second aspect of portability is \textit{permanence}: whether user data is retained by a system when a user switches algorithms. Data accumulation is crucial for large-scale machine learning platforms such as recommender systems, and operators are likely to retain as much data as possible, even after users leave. However, countervailing principles exist: Article 5(1)(c) of the GDPR establishes a \textit{data minimization} principle, requiring data to be limited to what is necessary, and Article 5(1)(e) restricts retention to no longer than necessary. It could be argued that a departing user’s profile is no longer needed by the original platform, implying non-permanence.

\begin{figure}
    \centering
    \includegraphics[width=0.45\textwidth]{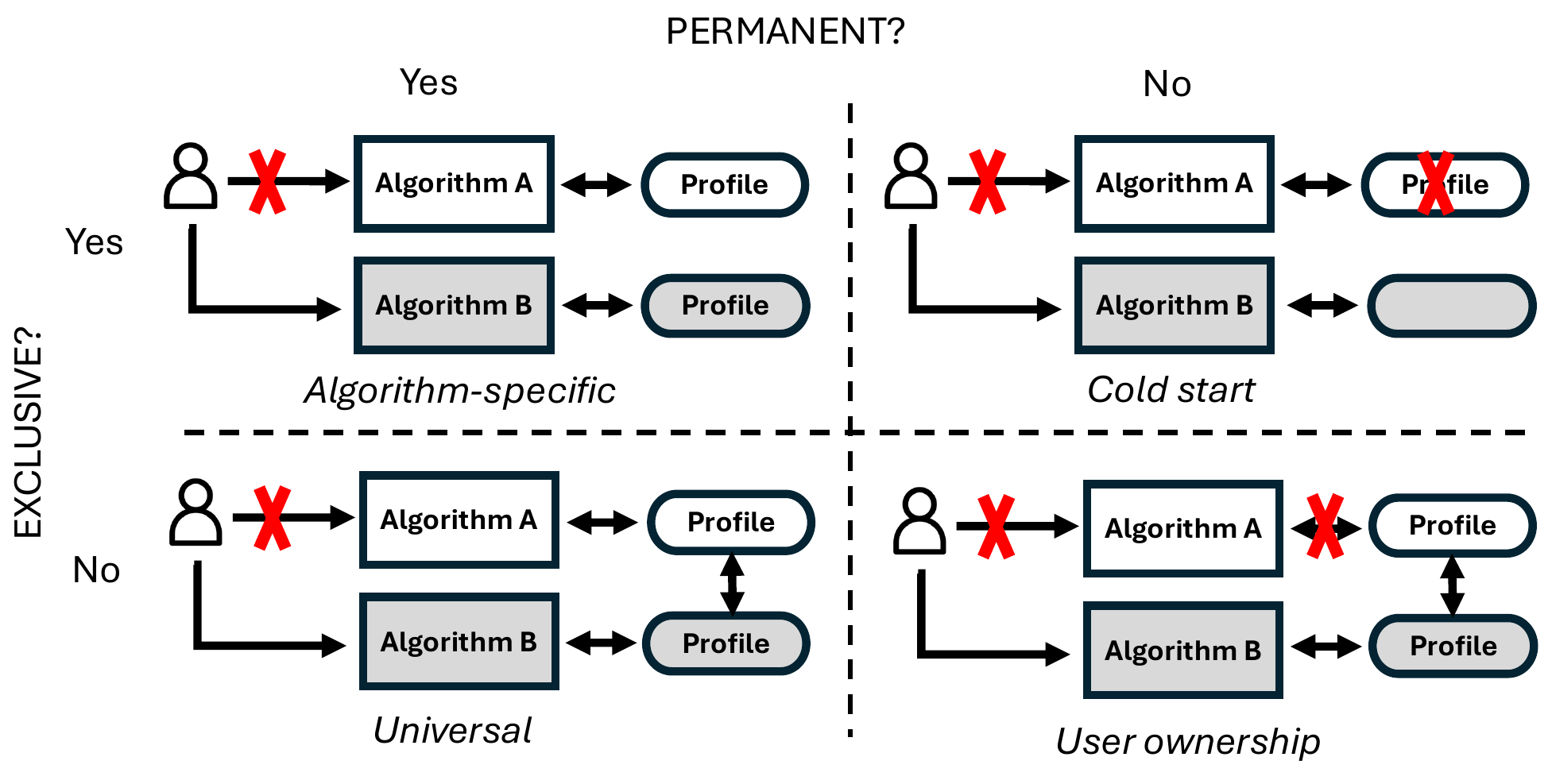}
    \caption{Profile portability options.}
    \label{fig:portability-options}
    \Description{Diagram titled “Profile portability options” showing four scenarios in a 2×2 grid. Top row labeled “Exclusive? Yes,” bottom row “No.” Columns show different portability designs: Algorithm-specific (top-left), Cold start (top-right), Universal (bottom-left), and User ownership (bottom-right).

    Each panel includes a user icon, two boxes labeled “Algorithm A” and “Algorithm B,” and one or more “Profile” shapes connected by arrows. In Algorithm-specific, each algorithm has its own separate profile. In Cold start, switching breaks the connection to the existing profile (marked with a red X) and creates a new empty profile. In Universal, both algorithms share a single common profile. In User ownership, the profile moves with the user from one algorithm to another, with the previous connection removed (red X) and a new one established.}
\end{figure}

The interaction between exclusivity and permanence gives us four different conditions as depicted in Figure~\ref{fig:portability-options}.\footnote{Icons courtesy of The Noun Project.}

\textbf{Algorithm-Specific} (Exclusive/Permanent): This reflects a competitive, unregulated market where users lack control over their profiles. Algorithms retain data when users switch and do not allow it to be transferred to competitors.

\textbf{Cold Start} (Exclusive/Non-permanent): This is the most restrictive case: users cannot transfer data, and profiles are deleted upon switching, making every user a cold-start case.

\textbf{User Ownership} (Non-exclusive/Non-permanent): Users carry their profiles across algorithms, creating blended histories, while platforms cannot access data of non-subscribers. Users effectively own and transfer their data.

\textbf{Universal Profile} (Non-exclusive/Permanent): Users bring their data, but platforms retain access even after users leave, resembling a shared profile database. This may arise in shared infrastructures \cite{hogg2024shapingfuturesocialmedia}.

These designs are not hypothetical. \textit{Algorithm-Specific} is standard, with platforms owning training data. \textit{Cold Start} appears in systems that allow data deletion, studied as \textit{machine unlearning} \cite{bourtoule2021machine}. \textit{Universal} designs occur in shared-protocol systems or where multiple algorithms operate on one platform. \textit{User Ownership} exists in ecosystems enabling cross-platform data transfer, such as importing music into Spotify via third-party tools\footnote{\url{https://www.spotify.com/us/import-music/}}


These conditions give rise to research questions that we explore in this work. Following the example of \cite{buhayh2024decoupled}, we concentrate on the simplified case where there are only two recommenders: one serving the general population -- we call this the Generic recommender $R^G$ -- and a second Niche recommender system $R^N$ that has been created to meet the needs of a subset of users for whom the generic one does not serve well.

\textbf{RQ1}: How does the \textit{exclusivity} of a profile impact outcomes for consumers transitioning from one recommender to another? We might expect that having a profile tied to a particular recommender would mean that consumers who switch will receive lower-quality recommendations. 

\textbf{RQ2}: How does the \textit{permanence} of a profile impact consumers who are not changing recommenders? If profiles are not permanent, consumers (e.g., Algorithm A consumers in the User Ownership case) receive recommendations generated from a smaller, potentially less diverse dataset after Niche consumers have left. 

\textbf{RQ3}: How do these profile options impact the utility for item providers? For example, we might expect consumers to switch more in the Cold Start condition, as it is more difficult for a recommender to build an effective profile. This could benefit providers with broader portfolios since Niche consumers would spend more time with the Generic recommender $R^G$. 

\section{Related Work}
\subsection{Multistakeholder fairness in recommender systems} 
There is much existing research examining and addressing fairness and disparities among different groups of stakeholders \cite{Abdollahpouri2019, RanjbarKermany2021, Smith2023, Wu2022}. The multistakeholder recommendation concept categorizes stakeholders on these platforms into three main groups: \(consumers\), who consume recommendations; \(providers\), who produce the items that are recommended; and the \(system\), which is made up of the datasets, recommendation algorithm, and platform operating it \cite{burke2017multisided, Abdollahpouri2020}. \citet{ekstrand2022fairness} found that some individuals and groups are treated unfairly on these platforms, leading to disparities in utility across stakeholders and among stakeholder groups. As noted in the studies of multisided platforms \cite{evans2016matchmakers}, such systems are sustainable only if they provide utility to all parties involved. 

Additional research shows that recommender systems often treat providers disparately, primarily due to popularity bias inherent in collaborative filtering algorithms \cite{Abdollahpouri2019a, Abdollahpouri2019b}. These biases can make it harder for new providers to do well \cite{Gope2017}, which impacts the inclusivity and diversity of recommender ecosystems. Many businesses and individuals rely on platforms that use such recommender systems for income \cite{Dalal2023, AlvarezdelaVega2021}, and so the fair treatment of providers in such systems is an important topic.

Furthermore, simulation studies suggest that unfair treatment of providers -- particularly niche providers -- incentivizes providers to adapt their content to align with what the recommender will recommend \cite{hron2023modeling, yao2023bad}. This theoretical finding concurs with user studies of providers who report prioritizing what they believe the algorithm favors over their true creative intentions \cite{Choi2023-uk}.

\subsection{Algorithmic pluralism and middleware}

\citet{jain2024algorithmic} advocate algorithmic pluralism — individuals with different characteristics should have an equal opportunity for outcomes — through Joseph Fishkin’s theory of bottlenecks, focusing on decision points that shape individual opportunities. They highlight alleviating severe bottlenecks — the mechanisms that create or constrain opportunities — while broadening access to opportunities for diverse users. This work provides a framework for reasoning about equality of opportunity in recommender systems. Extending prior work on how design choices cascade into outcomes \cite{sambasivan2021everyone, boyd2021datasheets, ghai2022cascaded}, \citet{jain2024algorithmic} center external societal factors shaping algorithmic opportunity structures.

A complementary response to the homogeneity of users' experience is \textit{middleware}, introduced by \citet{fukuyama2021save} and expanded in \cite{hogg2024shapingfuturesocialmedia, singh2024reimagining}. Middleware lets users choose among alternative algorithms instead of platform-defined ones. Developers can design algorithms for different user groups, and users can select those that match their needs. Similarly, \citet{rajendra2023three} envision a “friendly neighborhood algorithm store” for customizing social media feeds. Yet prior work warns that such flexibility can worsen filter bubbles and may need platform oversight \cite{Fukuyama2021-my, Ovadya_undated-ch}.

Building on these ideas, \citet{buhayh2024decoupled} studied algorithmic choice in a decoupled recommender system architecture that allows multiple algorithms to coexist, including those tailored to specific niche audiences. This work found that enabling users to select recommender algorithms that maximize their utility also improved the utility of the creators surfaced by those recommender systems. This work, extended in \cite{buhayh2025simulating}, shows that algorithmic choice can increase utility for both consumers and niche providers, highlighting the potential benefits of pluralistic recommender ecosystems.

\subsection{Simulation-based evaluation}

A central focus in recommender systems research is evaluating outcomes. Researchers assess how variables affect user utility using metrics such as accuracy, fairness, serendipity, and diversity \cite{zangerle2022evaluating}. Offline experiments, based on observed data, are the primary way to test performance before real-world deployment, allowing low-stakes evaluation. Industry experiments often use A/B testing, comparing user groups to measure statistically significant outcome differences in live settings \cite{hofmann2016online}. Laboratory studies also assess user satisfaction and collect feedback \cite{kelly2009methods}.

Simulation is the process of using observed or synthetic data to conduct experiments that resemble real-world scenarios, based on assumptions about user behavior. Simulation methods bridge offline and online evaluation by creating controlled ecosystems for testing recommender systems under specific assumptions. The need for simulation strategies stems from the limitations of static data and offline evaluation \cite{kouki2020lab}, as popular offline metrics are too strict and reward only user-item clicks or purchases, ignoring the multifaceted nature of user interactions. On the other hand, online evaluations are typically inaccessible to researchers outside of industry due to the need for a large user base and their high-stakes nature. 
 
 
Simulation research has long been used to support and investigate hypotheses in information retrieval and recommender systems and has been used to study long‑term dynamics and ecosystem impacts \cite{Yao2021SimulatedUsers, yao2023bad, Jannach2015Recommend, Hosanagar2009Sales, hazrati2022recommender, hron2023modeling, merinov2024positive}, user choice and interaction models \cite{szlavik2011diversity, rahdari2022ranked}, and conversational recommender systems \cite{balog2023user}. Simulation environments that have been tested and used by recommender systems researchers include RecSim~NG~\citep{mladenov2021recsimng}, Siren \citep{bountouridis2019siren}, RecoGym \citep{rohde2018recogym}, RecLab \citep{krauth2020offline}, and SMORES \cite{buhayh2025simulating}. However, SMORES is the only such system that directly supports multiple competing recommendation algorithms and simulates user choice among them. 

\subsection{Data governance and profile portability}

Previous work on profile portability focuses on technical feasibility. \citet{Abel2011-he} demonstrate that profile portability strategies significantly enhance personalization while also mitigating the severity of the cold-start problem . \citet{wischenbart2012user} show that creating a single unified profile across social media platforms is feasible using schema extraction of existing profiles. \citet{Bouraga2016-mj} propose a portable unified profile, a single profile which can be filled out once and used in multiple services – somewhat similar to the Universal condition that we explore. \citet{heitmann2010architecture} propose an architecture for privacy-enabled user profile portability, enabling users to manage their data for recommendations across contexts. In a study in which participants were asked to compare two social media profile portability systems (a direct transfer or one where they had access to their data), \citet{jamieson2023escaping} concluded that for portability to be empowering, it must be context-specific.

Data access and portability have been addressed in recent legislation. The EU’s GDPR (2016) established portability as a user right, followed by the Digital Markets Act (2022) and Data Act (2023), while the CCPA (2018, strengthened in 2021) introduced similar rights in California; comparable U.S. federal proposals remain unenacted, and Japan’s APPI revision (2020) grants related access rights \cite{chao2025data, copra2021, access2021, nakashima2022comparison}.

Industry has also emphasized interoperability, with policy reports advocating real-time data portability and initiatives such as the Data Transfer Project (now the Data Transfer Initiative) enabling direct data transfer between platforms, though adoption remains limited \cite{schweitzer2019competition, willard2018intro}.

Importantly, data portability (\textit{non-permanence}) as outlined in Article 20 of the GDPR does not necessarily interact with the right to be forgotten (Article 17); a data portability request "does not automatically trigger the erasure of the data from the systems." \cite{wp2017dataportability} Nonetheless, data portability allows for data \textit{non-exclusivity}. As separate rights, these could be applied together, separately.

%% file: 2_smores.tex
\section{SMORES}

The SMORES (Simulation Model for Recommender EcoSystems) framework~\cite{buhayh2024decoupled} is a simulation environment for modeling multistakeholder recommender ecosystems. SMORES enables the study of complex interactions among consumers, providers, and recommendation algorithms under controlled conditions, making it suitable for evaluating ecosystem-level design questions such as profile portability and algorithmic choice. It is an open-source system written in Python and available from GitHub. \footnote{https://github.com/that-recsys-lab/smores}

SMORES represents a recommender ecosystem with three primary stakeholders: \textit{consumers} ($j \in J$), who receive and act on recommendations; \textit{providers} ($v \in V$), who produce items ($i \in I$); and \textit{recommenders} ($k \in K$), which generate recommendation slates. Consumers interact with a single recommender at a time, selecting items from slates based on a customizable preference function. In our experiments, we model user preferences with normalized genre-based preference vectors built from real-world datasets. Providers gain utility from consumers selecting their items from recommendation slates, while recommenders collect interaction data to refine recommendations. The framework supports dynamic consumer behavior, allowing consumers to switch between recommenders based on their satisfaction/utility, and tracks utility outcomes over time. By simulating these interactions, SMORES facilitates the evaluation of utility for diverse stakeholder groups, particularly underserved niche consumers and providers.

%% file: 3_methodology.tex
\section{Methodology}

This study extends the SMORES framework to investigate how profile portability policies within algorithmic choice impact utility outcomes for \textit{Niche} and \textit{Generic} consumers and \textit{Niche} and \textit{Generic} providers in a recommender ecosystem. 

\subsection{Datasets}

We chose three datasets on which to base our simulations: Amazon Video Games Sales (based on the larger Amazon Reviews 2023 dataset \cite{hou2024bridging}), Goodreads \cite{wan2018item}, and MovieLens 10M \cite{harper2015movielens}. We selected datasets in which it is reasonable to assume that a consumer would choose a single item per day, and where consumers have a low tendency to interact with the same item again, unlike music or short videos, where a consumer might interact with the same item multiple times.

We used k-core filtering on each dataset to maintain a similar size (see Table~\ref{tab:dataset_stats}). We augmented the MovieLens dataset with TMDb data for provider (movie studio) details \footnote{https://www.themoviedb.org/}, and the Amazon Video games with RAWG data for provider (game developer) and genre details \footnote{https://rawg.io/}. We then filtered out all items without a provider match to ensure data completeness.

To study the impact of profile portability, we simplified the experiments by defining one genre per dataset as a niche genre and labeling the remaining genres as \textit{Generic} – as in the work of \cite{buhayh2024decoupled, buhayh2025simulating}. To select niche genres, we defined supply as the geometric mean of provider and item counts for each genre and measured the mismatch between consumer demand and supply. We restricted our attention to genres between the 30th and 80th percentiles to avoid low demand and very popular genres. Within this set, we selected the genre with the largest mismatch and labeled it as \textit{Niche}. We then labeled each provider by the dominant genre among the items they offer, and labeled them as \textit{Niche} if their most-produced genre is niche. For consumers, we defined them as \textit{Niche} if their highest-rated genre is a niche genre. See Table~\ref{tab:dataset_stats}.

\begin{table*}[t]
\centering
\small
\begin{tabular}{|l|c|c||c|c||c|c||c|c|}
\hline
\multicolumn{3}{|c||}{Dataset}
& \multicolumn{2}{c||}{Items}
& \multicolumn{2}{c||}{Providers}
& \multicolumn{2}{c|}{Consumers} \\
\cline{1-9}
Name
& K-core
& Niche genre
& Total
& Niche
& Total
& Niche
& Total
& Niche \\
\hline
MovieLens 10M & 30 & Romance & 5155 & 893 (17.32\%) & 1950 & 18 (0.92\%) & 47899 & 329 (0.69\%) \\
Goodreads & 30 & Fantasy--Paranormal  & 8547 & 2799 (32.75\%)  & 4255 & 748 (17.58\%)  & 7723 & 523 (6.77\%) \\
Amazon Video Games & 5 & Platformer & 3726 & 225 (6.04\%) & 1526 & 70 (4.59\%) & 16160 & 321 (1.99\%) \\
\hline
\end{tabular}
\caption{Dataset statistics with niche breakdowns.}
\bumpup
\label{tab:dataset_stats}
\end{table*}

\subsection{Experimental setup}
\label{sec:setup}
We simulate an ecosystem with two recommenders \footnote{We expect that recommenders for different audiences would emerge organically, rather than being predefined as they are here, and plan to study such emergence in future work.}: a \textit{Generic} recommender $R^G$ serving all content genres and a \textit{Niche} recommender $R^N$ specializing in the niche genre. The two recommenders differ in their cold start recommendations and the items they recommend. The niche recommender only recommends items where the niche genre is present in the item feature vector. In the example of MovieLens dataset, the Romance genre must be present in the movie’s genres for the item to be recommended by the niche recommender. Whereas the generic recommender recommends from all genres, including the niche genres.

All consumers are initially connected to $R^G$. To allow interaction histories to form and to avoid cold-start noise, switching is disabled for the first two cycles. After this warm-up period, the decision model takes effect, and consumers may switch among recommenders based on the utility they experience; they will consider switching only if their utility falls below the threshold $\tau$ = 0.2. We also run the simulation with $R^G$ as the only recommender to provide a baseline (non-switching) condition. Although comparing different algorithm decision models is beyond the scope of this work, \citet{buhayh2025simulating} explored various models and observed similar switching patterns across models. We therefore use the simple threshold model here.

The simulation process is captured in pseudocode below. Note that a consumer $j$ has an "attached" recommender, $j_k$, the recommender they currently receive recommendations from. Let $\hat{K}$ be the set of active recommenders , and let $V$ denote the set of providers.


\begin{algorithm}
\caption{SMORES Simulation Process}\label{alg:smores}
\ForEach{cycle}{
\ForEach{$k \in \hat{K}$}{
train($k$);
}
\ForEach{day $t$}{
\ForEach{consumer $j$}{
$recs \gets j_k.\mathrm{recommendations}(j)$\\
$\bar{\mu}_{j,k,t} \gets \mathrm{update\_consumer\_utility}(j, recs)$\\
$i \gets \mathrm{select\_item}(j, recs)$\\
$\mu_V \gets \mathrm{update\_provider\_utility}(i, recs)$\\
\If{$\bar{\mu}_{j,k,t} < \tau$}{
\ForEach{$k' \in \hat{K},~k' \neq j_k$}{
\If{$\bar{\mu}_{j,k',t} \ge \bar{\mu}_{j,k,t}$}{
$j_k \gets k'$
}
}
}
}
}
\ForEach{consumer $j$}{
$\mathrm{manage\_profile}(j)$
}
}
\end{algorithm}

The \texttt{update\_consumer\_utility} function computes consumer utility as the relevance of a recommendation list to consumer $j$'s genre preferences. For each recommended item $i \in \ell_{j,k}$, instantaneous utility is $\mu_{j,i} = f_i \cdot p_j$, where $f_i$ is the item's feature vector and $p_j$ is the consumer preference vector. Scores are normalized by feature-set size, converted to selection probabilities with a softmax as in \cite{buhayh2025simulating}, and incorporated into the running utility estimate with recency bias $\beta$: $\bar{\mu}_{j,k,t} = (\bar{\mu}_{j,k,t-1} \times \beta + \mu_{j,k,t}) / (1 + \beta)$. We set $\beta = 2.0$.

The \texttt{update\_provider\_utility} function updates provider utility, which is defined by the number of clicks on an affiliated item by a consumer, a common e-commerce metric. Given the item that was selected and the recommendation list that item appeared in, utility is updated for the provider of that item.

Based on the updated consumer utility $\bar{\mu}_{j,k,t}$, the consumer decides whether to stay with the current recommender or switch to another. We model user decision-making using a simple threshold-based approach; if consumer utility is below the threshold, and the utility for another recommender $k'$ is greater, the consumer will switch to the $k'$ recommender at the end of the cycle.

The \texttt{manage\_profile} function applies the active portability policy after a consumer switches recommenders. In \textit{Cold Start}, the old profile is deleted. In \textit{User Ownership}, the profile is transferred to the new recommender and removed from the old one. In \textit{Universal Profile}, recommenders share access to the same profile. In \textit{Algorithm-Specific}, each recommender keeps its own private profile and never transfers it. A user profile consists of the consumer's interaction history: clicks.

Each experiment runs for 10 cycles of 3 days, with each consumer receiving 5 recommendations per day. The first four items come from the ranking algorithm and the fifth from popularity-based sampling to introduce exploration and expand the interaction set. Clicks are generated from the dot product between consumer preferences and item genre vectors, as in prior SMORES work \cite{buhayh2024decoupled}, and are then used to train the recommender. To limit repeated exposure, an item shown to a user three times without a click is withheld for the next cycle.

\subsection{Algorithms and Candidate Generation}

To provide a broader view of how different data portability conditions affect user utility, we evaluate three widely used ranking algorithms, implemented using the LensKit \footnote{https://lkpy.lenskit.org/stable/} and Implicit \footnote{https://benfred.github.io/implicit/} libraries: Alternating Least Squares \cite{hu2008collaborative}, Bayesian Personalized Ranking \cite{rendle2012bpr}, and Item-KNN \cite{deshpande2004item}.

We train the algorithms on the interactions that occur during the simulation to generate a slate of five items. Because more complex recommendation models require sufficient interaction history to compute recommendation slates, we implement two levels of fallback recommendations for the cold-start conditions that arise in the simulation. When there are insufficient consumers to make predictions, each recommender generates popularity-based cold-start predictions by conducting weighted sampling from the item catalog. When a new consumer joins the recommender and has no associated profile, the recommender operates as a non-personalized popularity-based recommender, using data from its current consumers. For efficiency, we train recommenders only once per cycle.

\subsection{Evaluation Protocol}

We report stakeholder utilities as the primary outcomes of our experiments. Our analysis examines how different data portability scenarios affect stakeholder utility in recommender systems, across multiple datasets and algorithms.

For each dataset and recommendation algorithm, we first ran the simulation using only the \textit{Generic} recommender to establish a baseline. We then repeated the simulations across the three datasets and three algorithms for each of the four profile portability conditions, using five different seeds per scenario. 

We focus on the utility achieved in the last 5 simulation cycles when the distribution of consumers has (more or less) achieved a steady state. For both consumers and providers, we average the utility over the last five cycles, then average across types (\textit{Niche} or \textit{Generic}) when reporting the results. The trajectories of both consumers and providers across the simulation time span are of interest but are omitted here due to space constraints.

\subsection{Reproducibility Details}
The experiment configuration and pre-processed datasets are maintained in an experiment repository managed using Data Version Control (DVC). \footnote{\url{https://github.com/that-recsys-lab/smores_umap_2026}}. 

%% file: 4_results.tex
\section{Results}

The results of our experiments show how profile exclusivity and permanence shape utility for both consumers and providers across recommenders, algorithms, and datasets (see Tables~\ref{tab:consumer-utilities} and \ref{tab:provider-utility}). Our results are consistent with the prior work of \citet{buhayh2024decoupled,buhayh2025simulating}, showing that adding a specialized recommender yields higher utility for niche providers and consumers. In the study of the effects on consumers of profile portability (RQ1) and profile permanence (RQ2), and the effects on providers (RQ3), we find that generally algorithm-specific and cold-start scenarios yield the highest utility for niche consumers relative to the baseline condition, with generic consumers gaining less utility than their \textit{Niche} counterparts. The patterns for providers are less consistent: \textit{Niche} providers achieve a high utility under most conditions, while \textit{Generic} providers lose some utility in almost all cases due to the introduction of the niche recommender.


\subsection{Consumer Utility}

As shown in Figure \ref{fig:consumer-utility} and Table~\ref{tab:consumer-utilities}, when a niche recommender is introduced, niche consumer utility increases relative to the baseline across most datasets (35/36 scenarios with positive change) with changes ranging from -4.2\%--192.2\% in mean utility. Generic consumers' changes are more likely to be negative but smaller in magnitude, ranging from -26.4\%--3.9\% (7/36 scenarios with positive change) across datasets when the niche recommender is introduced.

\textit{Dataset comparison}:
Niche Amazon Video Games consumers show the largest increase, ranging from +22.2\%--192.2\% relative to baseline, compared with MovieLens and Goodreads, which exhibit smaller, and in some cases negative, changes of -4.2\%--46.5\% and +1.0\%--7.1\%, respectively. The smaller change in MovieLens and Goodreads could be attributed to the smaller gap in the arithmetic means for niche and generic users in the baseline condition.


\textit{Algorithm comparison}: 
Consumer utility also varies across algorithms, as shown in \ref{fig:consumer-utility}. There is no noticeable difference in utility changes across algorithms using the Goodreads dataset, which may be attributed to the high genre correlation among items; the average pointwise similarity is 0.0825, compared with 0.0384 for MovieLens and 0.0235 for Amazon Video Games. As a result, the niche genre is recommended at similar rates across conditions, accounting for 32.75\% of items in the Goodreads dataset, compared to 17.32\% for MovieLens and 6.04\% for Amazon Video Games. Interestingly, ItemKNN provides the highest increase in utility for niche consumers across all datasets in the Algorithm-Specific and Cold start conditions, but under User Ownership in the MovieLens dataset, niche consumers lose a small amount of utility. Generic consumers, on the other hand, lose most utility under ItemKNN. This could be because the migration of niche users away from the generic recommender has a bigger impact on the diversity of recommendations that the ItemKNN recommender can provide. 


\textit{Portability scenario comparison}:
Across the four profile portability conditions, niche consumers generally derive the largest gains in the \textit{exclusive} conditions (algorithm-specific and cold-start scenarios +2.6\%--192.2\%), especially in the MovieLens and Amazon Video Games datasets. Generic consumers’ utility typically changes within a small range close to the baseline value. 


\input{consumer_results_fig}

\begin{table*}[t]
\centering
\small
\begin{tabular}{llllllll}
\toprule
 &  & \multicolumn{2}{c}{MovieLens 10M} & \multicolumn{2}{c}{Goodreads} & \multicolumn{2}{c}{Amazon Video Games} \\
Condition & Algorithm &  Niche & Generic & Niche & Generic & Niche & Generic  \\
\midrule
\multirow[t]{3}{*}{Baseline} & BPR & 0.122 & 0.111 & 0.178 & 0.211 & 0.097 & 0.141 \\
 & ALS & 0.138 & 0.120 & 0.181 & 0.228 & 0.100 & 0.173 \\
 & ItemKNN & 0.146 & 0.122 & 0.168 & 0.193 & 0.104 & 0.174 \\
\cline{1-8}
\multirow[t]{3}{*}{Algo-specific} & BPR & 0.136 (+12.1\%) & 0.112 (+0.4\%) & 0.183 (+2.6\%) & 0.209 (-0.7\%) & 0.143 (+47.5\%) & 0.137 (-3.1\%) \\
 & ALS & 0.142 (+2.8\%) & 0.119 (-0.4\%) & 0.188 (+3.6\%) & 0.227 (-0.4\%) & 0.147 (+46.8\%) & 0.170 (-1.9\%) \\
 & ItemKNN & 0.214 (+46.5\%) & 0.119 (-2.4\%) & 0.180 (+7.1\%) & 0.193 (-0.2\%) & 0.303 (+192.2\%) & 0.168 (-3.4\%) \\
\cline{1-8}
\multirow[t]{3}{*}{Cold start} & BPR & 0.140 (+15.2\%) & 0.110 (-1.2\%) & 0.180 (+1.0\%) & 0.201 (-4.3\%) & 0.152 (+57.0\%) & 0.135 (-3.9\%) \\
 & ALS & 0.149 (+7.3\%) & 0.115 (-3.7\%) & 0.186 (+2.3\%) & 0.225 (-1.0\%) & 0.152 (+52.3\%) & 0.165 (-5.0\%) \\
 & ItemKNN & 0.214 (+46.5\%) & 0.111 (-8.8\%) & 0.179 (+6.0\%) & 0.195 (+1.0\%) & 0.303 (+192.2\%) & 0.143 (-17.5\%) \\
\cline{1-8}
\multirow[t]{3}{*}{User ownership} & BPR & 0.136 (+11.4\%) & 0.114 (+2.3\%) & 0.187 (+5.0\%) & 0.205 (-2.7\%) & 0.126 (+30.2\%) & 0.147 (+3.9\%) \\
 & ALS & 0.149 (+7.5\%) & 0.120 (+0.7\%) & 0.186 (+2.8\%) & 0.224 (-1.4\%) & 0.127 (+26.9\%) & 0.166 (-4.3\%) \\
 & ItemKNN & 0.140 (-4.2\%) & 0.113 (-7.7\%) & 0.179 (+6.4\%) & 0.194 (+0.2\%) & 0.133 (+28.2\%) & 0.128 (-26.4\%) \\
\cline{1-8}
\multirow[t]{3}{*}{Universal} & BPR & 0.135 (+11.0\%) & 0.113 (+1.6\%) & 0.182 (+2.0\%) & 0.206 (-2.1\%) & 0.123 (+27.5\%) & 0.133 (-5.9\%) \\
 & ALS & 0.146 (+5.6\%) & 0.119 (-0.2\%) & 0.185 (+2.0\%) & 0.223 (-2.1\%) & 0.125 (+24.9\%) & 0.157 (-9.3\%) \\
 & ItemKNN & 0.157 (+7.1\%) & 0.116 (-4.8\%) & 0.180 (+6.7\%) & 0.193 (-0.3\%) & 0.127 (+22.2\%) & 0.152 (-12.5\%) \\
\cline{1-8}
\bottomrule
\end{tabular}
\caption{Mean consumer utility for different recommendation algorithms, datasets, and portability conditions. Percent difference from baseline shown in parentheses.}
\bumpup
\label{tab:consumer-utilities}
\end{table*}

\subsection{Provider Utility}

As shown in Figure~\ref{fig:provider-utility} and Table~\ref{tab:provider-utility}, niche providers generally achieve higher utility than the baseline (single generic recommender) with mean utility changes ranging from -63.2\%--356.9\%, and generic providers see their utility dip below the baseline in all but two scenarios, ranging from -34.7\%--2.9\%. This is consistent with the results reported in \cite{buhayh2024decoupled,buhayh2025simulating}, which also show that niche providers usually achieve higher utility when a specialized recommender is introduced, while utility for generic providers decreases or remains about the same. 

\textit{Dataset comparison}:
An important distinction in the Goodreads dataset is that niche providers actually lose utility on average in all cases except two scenarios. However, niche providers in the Goodreads dataset perform relatively well even in the baseline condition (mean utility of 1.622--4.124 compared to the generic providers 2.182--3.950), whereas MovieLens and Amazon Video Games niche providers start from significantly lower baseline utility in comparison to their generic counterparts ( 1.576--4.420 mean utility for niche and 6.822-7.759 for generic, and 1.486--1.697 mean utility for niche and 7.546--9.305 for generic, respectively). A dominant factor in these differences is the number of niche providers and items in each dataset: niche items account for 32.75\% of providers’ catalogs in Goodreads vs. 17.32\% in MovieLens and 6.04\% in Amazon Video Games.


\textit{Algorithm comparison}: 
Similarly to niche consumer utility, ItemKNN yields the highest niche provider utility across most datasets and conditions. In the MovieLens and Amazon Video Games datasets, the mean utility for niche providers under ItemKNN ranges from +11.1\%--356.9\% and from 18.6\%--166.7\%, respectively, compared to a range of -17.5\%-- -5.5\% in the Goodreads Dataset. The BPR algorithm had a majority-positive impact on niche providers (8/12 scenarios with positive change), whereas the ALS algorithm had a mostly positive impact on niche providers in Movielens and Video Games (7/9 scenarios with positive change), but only a negative impact on Goodreads niche providers. For both niche and generic providers, there are no clear patterns.


\textit{Profile scenario comparison}:
Similarly to the Algorithm analysis, we did not notice any patterns across the portability scenarios. For instance, the MovieLens dataset and the BPR algorithm, niche providers gain the most in utility in the Algorithm-Specific and User ownership scenarios, which are fully opposite in terms of exclusivity and permanence. For Goodreads, the only niche providers who gain utility are those under the BPR algorithm in the User Ownership and Universal scenarios (non-exclusive). And in the Amazon Video Games dataset, the greatest gains in utility are under the Universal condition with the BPR algorithm or under the Cold start scenario under the ItemKNN algorithm, which are also fully opposite in terms of exclusivity and permanence. For generic providers, there is greater similarity in the changes in utility, but no clear winner.

\input{provider_results_fig}

\begin{table*}[t]
\centering
\small
\begin{tabular}{llllllll}
\toprule
 &  & \multicolumn{2}{c}{MovieLens 10M} & \multicolumn{2}{c}{Goodreads} & \multicolumn{2}{c}{Amazon Video Games} \\
Condition & Algorithm &  Niche & Generic & Niche & Generic & Niche & Generic  \\
\midrule
\multirow[t]{3}{*}{Baseline} & BPR & 4.420 & 7.759 & 4.124 & 3.950 & 1.486 & 9.305 \\
 & ALS & 1.576 & 6.822 & 1.622 & 2.308 & 1.610 & 7.828 \\
 & ItemKNN & 3.049 & 6.984 & 2.080 & 2.182 & 1.697 & 7.546 \\
\cline{1-8}
\multirow[t]{3}{*}{Algo-specific} & BPR & 13.836 (+213.0\%) & 5.704 (-26.5\%) & 2.555 (-38.0\%) & 3.983 (+0.8\%) & 2.037 (+37.1\%) & 8.769 (-5.8\%) \\
 & ALS & 2.443 (+55.0\%) & 6.192 (-9.2\%) & 1.401 (-13.6\%) & 2.176 (-5.7\%) & 1.877 (+16.6\%) & 7.241 (-7.5\%) \\
 & ItemKNN & 7.365 (+141.6\%) & 6.212 (-11.0\%) & 1.965 (-5.5\%) & 1.858 (-14.8\%) & 4.133 (+143.6\%) & 6.870 (-9.0\%) \\
\cline{1-8}
\multirow[t]{3}{*}{Cold start} & BPR & 3.941 (-10.8\%) & 5.067 (-34.7\%) & 2.408 (-41.6\%) & 2.687 (-32.0\%) & 2.677 (+80.2\%) & 7.753 (-16.7\%) \\
 & ALS & 3.170 (+101.2\%) & 5.678 (-16.8\%) & 1.441 (-11.1\%) & 2.197 (-4.8\%) & 2.284 (+41.9\%) & 7.438 (-5.0\%) \\
 & ItemKNN & 7.617 (+149.8\%) & 5.336 (-23.6\%) & 1.826 (-12.2\%) & 1.946 (-10.8\%) & 4.526 (+166.7\%) & 7.174 (-4.9\%) \\
\cline{1-8}
\multirow[t]{3}{*}{User ownership} & BPR & 12.419 (+180.9\%) & 7.594 (-2.1\%) & 5.199 (+26.1\%) & 3.722 (-5.8\%) & 3.072 (+106.7\%) & 9.578 (+2.9\%) \\
 & ALS & 1.707 (+8.3\%) & 5.803 (-14.9\%) & 1.475 (-9.0\%) & 2.156 (-6.6\%) & 1.582 (-1.7\%) & 6.846 (-12.5\%) \\
 & ItemKNN & 3.388 (+11.1\%) & 5.293 (-24.2\%) & 1.825 (-12.2\%) & 1.938 (-11.2\%) & 4.200 (+147.5\%) & 6.059 (-19.7\%) \\
\cline{1-8}
\multirow[t]{3}{*}{Universal} & BPR & 1.625 (-63.2\%) & 6.068 (-21.8\%) & 4.425 (+7.3\%) & 3.431 (-13.1\%) & 4.923 (+231.3\%) & 8.792 (-5.5\%) \\
 & ALS & 1.833 (+16.3\%) & 5.657 (-17.1\%) & 1.383 (-14.7\%) & 2.083 (-9.7\%) & 1.692 (+5.1\%) & 6.494 (-17.0\%) \\
 & ItemKNN & 13.930 (+356.9\%) & 4.759 (-31.9\%) & 1.716 (-17.5\%) & 1.816 (-16.7\%) & 2.012 (+18.6\%) & 6.082 (-19.4\%) \\
\cline{1-8}
\bottomrule
\end{tabular}
\caption{Mean provider utility for different algorithms and conditions and across the different datasets}
\bumpup
\label{tab:provider-utility}
\end{table*}

\section{Discussion}
In this work, we experiment with different profile portability conditions in a simulated environment to assess how these decisions affect stakeholder utility. We find that each condition involves trade-offs for consumers that must be carefully considered when implementing such systems. However, there were no consistent impacts for provider utility, although in most conditions niche provider utility was improved over the baseline, excluding the Goodreads dataset where their utility is already close to that of the generic providers. 

RQ1 asks about the impact of \textit{exclusivity} on consumers. The two exclusive conditions in our experiments are Algorithm-Specific and Cold Start. Our expectation was that the Algorithm-Specific and Cold Start conditions might result in lower utility for niche consumers since consumers are not able to bring data with them when switching recommenders. We did not find this to be the case and in fact, the Algorithm-Specific and Cold Start conditions resulted in more utility gain for niche consumers than the other conditions.

RQ2 asks about the impact of \textit{permanence} on consumers. The two permanent conditions in our experiments are Algorithm-Specific and Universal. Our expectation was that the non-permanent conditions -- Cold Start and User Ownership -- might result in lower utility for consumers, because consumers who stay will be served recommendations from a smaller, potentially less diverse dataset after niche consumers have left and taken their profiles with them (in the User Ownership case) or had them deleted (in the Cold Start case). We did not find this to be the case, as for generic consumers there was not a clear pattern in utility gain (or loss) across the conditions. However, the two biggest losses in utility gain for generic consumers occurred under the Cold Start and User Ownership conditions, with -17.5\% and -26.4\%, respectively, both under the Amazon Video Games dataset and the ItemKNN algorithm. 

RQ3 asks about the impact of the profile conditions generally on providers. Our expectation was that generic providers might benefit under the Cold Start condition (and niche providers might suffer) because more switching could happen, which could benefit providers with more diverse profiles. We also thought that the non-exclusive (Universal and User Ownership) conditions might be best for niche providers, because the niche recommender would have access to niche consumer profiles when they switch to the niche recommender. We did not find this to be the case, because there were not clear patterns across the portability condition scenarios. Niche providers sometimes did best under User Ownership or Universal scenarios, but in other datasets and algorithms, did the best under Algorithm-Specific or Cold Start.

In examining the results of these conditions, it appears that profile non-exclusivity and profile permanence can have a negative impact on the niche recommender by weakening its focus on its core niche mission. That is, when profiles are non-exclusive, an influx of curious, but ultimately uninterested users, brings with it a large number of user profiles that are unrelated to the narrow niche that the recommender is seeking to satisfy. This data amounts to noise relative to the task of modeling niche consumers. If profiles are permanent, those profiles stay with the niche recommender and continue to impact its recommendations. These effects are strongest with the ItemKNN algorithm, where recommendations are extrapolated from individual profiles. Provider utility is not as much effected because the niche recommender is still recommending only niche items although these might be segments of the item space with more mainstream appeal. 

This consideration also explains why the effect is much reduced in the Goodreads dataset. The niche users and generic users are not very different from each other, and therefore, the appearance of generic consumers in the niche recommender data has little impact.

If we think of non-exclusivity and non-permanence (that is, User Ownership) as the most sought-after profile portability properties, we find that this condition is not the one with the best utility in our simulations. Despite that, we find these results to be encouraging relative to profile portability generally. There are reasons other than recommendation accuracy to allow users to transfer and delete their data as desired, and these should be honored. Our work shows that it would be up the recommender systems themselves, especially those catering to a niche or targeted audience, to be judicious in the integration of data from other platforms and their retention of data when profile portability is supported. Our work supports efforts at data minimization by suggesting that wholesale data retention from migrating users is not always in the best interest of the recommender. 

%% file: consumer_results_fig.tex
\begin{figure}
    \centering
    \includegraphics[width=1\linewidth]{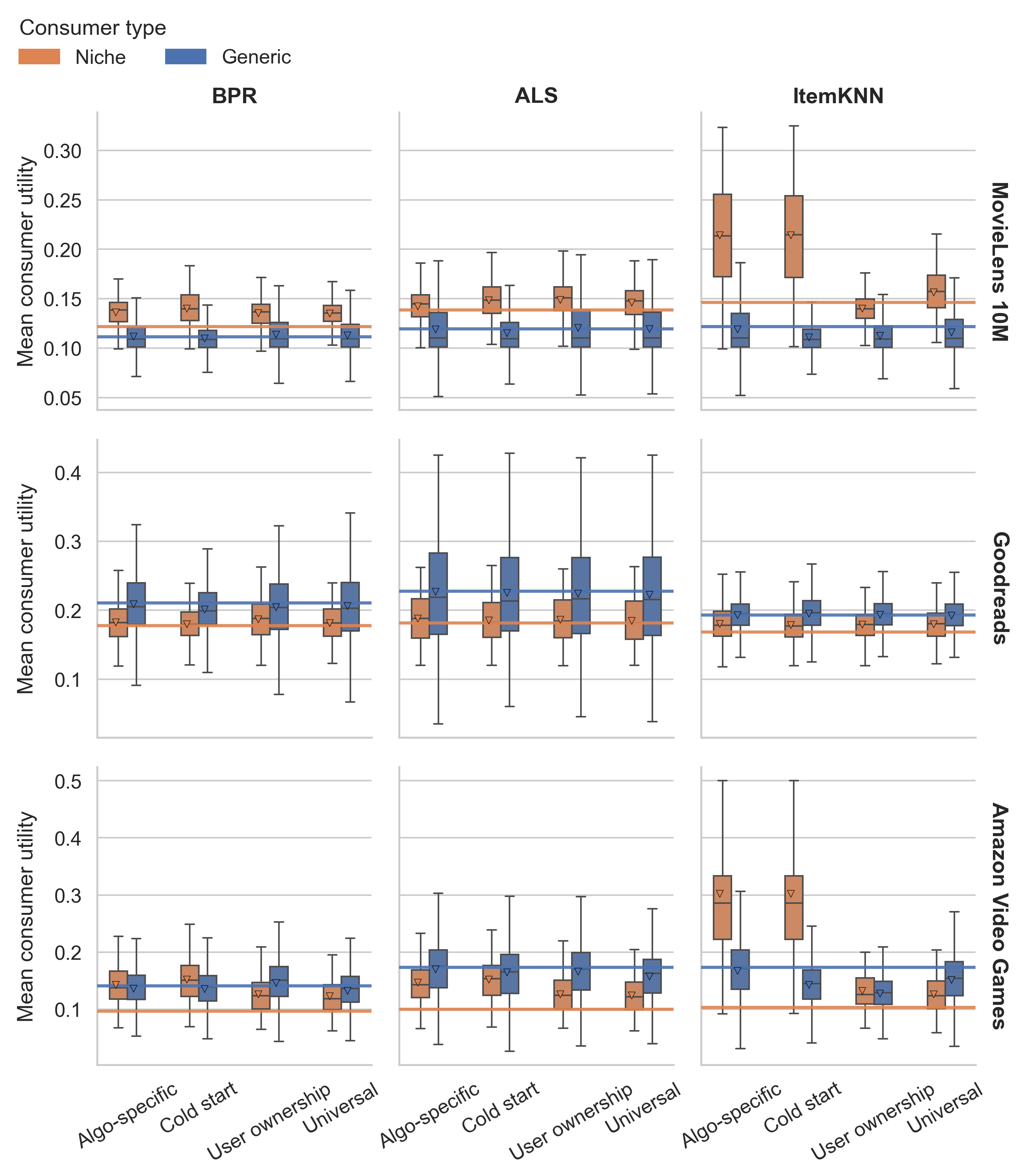}
    \caption{Platform portability impacts on mean consumer utility averaged over cycles 5 - 10. Rows are different datasets and columns are different algorithms. The horizontal lines represent the mean utility for the baseline condition, and the inverted triangles indicate the mean for the different conditions for consumers in that group.}
    \label{fig:consumer-utility}
    \Description{A 3×3 grid of boxplots showing mean consumer utility across three algorithms (BPR, ALS, ItemKNN; columns) and three datasets (MovieLens 10M, Goodreads, Amazon Video Games; rows). Each subplot compares four data portability conditions on the x-axis (Algorithm-specific, Cold start, User ownership, Universal). Boxplots are split by consumer type: orange for niche users and blue for generic users. Horizontal lines indicate average utility for each group.

    Across most panels, niche users tend to have slightly higher utility than generic users in MovieLens, especially under ItemKNN where niche utility is substantially higher. In Goodreads and Amazon datasets, generic users often have equal or higher utility than niche users, with larger variance in ALS. Differences across portability conditions are modest, though Cold start and Universal sometimes slightly reduce utility. Variability is higher in ALS and ItemKNN compared to BPR.}
\end{figure}

%% file: provider_results_fig.tex
\begin{figure}
    \centering
    \includegraphics[width=1\linewidth]{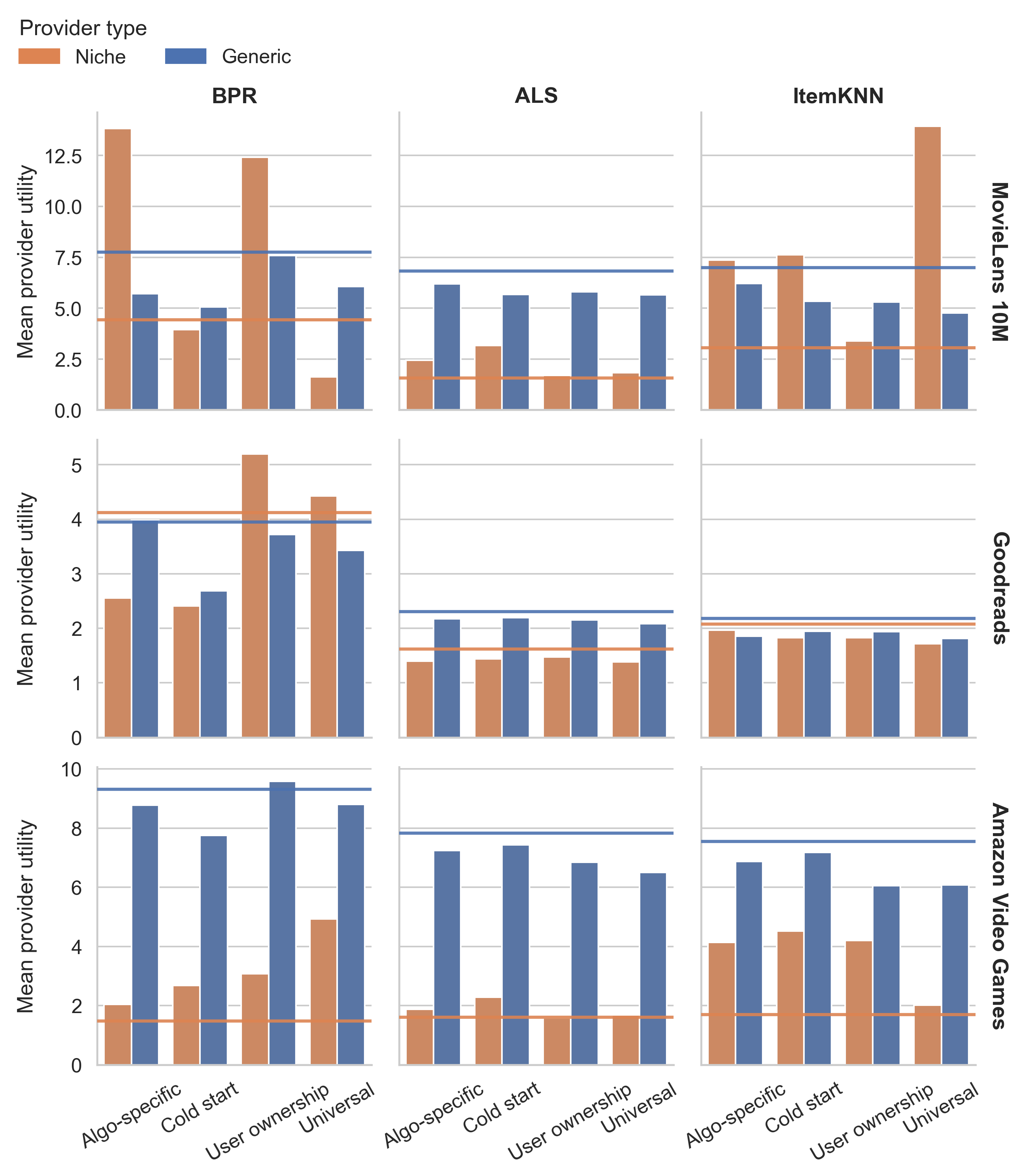}
    \caption{Algorithm sensitivity of mean provider utility for cycles 5 - 10. The horizontal lines represent the mean utility for the baseline condition, and the inverted triangles indicate the mean for the different conditions for providers in that group.}
    \label{fig:provider-utility}
    \Description{A 3×3 grid of bar charts showing mean provider utility across three algorithms (BPR, ALS, ItemKNN; columns) and three datasets (MovieLens 10M, Goodreads, Amazon Video Games; rows). Each subplot compares four portability conditions (Algorithm-specific, Cold start, User ownership, Universal). Bars are split by provider type: orange for niche providers and blue for generic providers. Horizontal lines indicate average utility for each group.

    In MovieLens, niche providers often outperform generic providers under BPR and ItemKNN, with especially large gains in Algorithm-specific and Universal settings. In contrast, generic providers dominate in ALS. In Goodreads, generic providers consistently have higher utility across all algorithms and conditions, though differences are smaller. In Amazon Video Games, generic providers strongly outperform niche providers across all algorithms, with large gaps in BPR and ALS. Portability conditions influence magnitude but not overall ranking patterns between provider types.}
\end{figure}

%% file: 5_conclusion.tex
\section{Conclusion}
Our results show that profile portability may not be a ``deal breaker'' for a recommendation middleware solution. Even in the most restrictive cases, where users cannot move their profile data, having algorithmic choice remains beneficial to recommendation consumers in most cases and almost always to providers, especially those targeting users distinct from majority. There may be no reason to wait for cooperation from platforms on portability questions before working towards recommendation ecosystems that support algorithmic pluralism. As noted above, data integration and retention policies maintained by recommendation platforms may allow them to avoid some of the negative effects noted here when portability is enabled. We leave the exploration of such policies for future work. 

Future work is also needed to explore the sensitivity of these findings to different experiment assumptions, including the choice of niche genre(s) and the model of recommendation uptake. We also believe that there are opportunities to explore user modeling of recommender switching behavior. We use a simple threshold model here, but note that a multi-arm bandit approach has also been successfully used in the SMORES model \cite{buhayh2025simulating}. A more complete model of a middleware ecosystem would also need to account for the utility of platforms themselves and modeling of the emergence of new recommenders in response to market demand. 